# Machine Learning–Driven Creep Law Discovery Across Alloy Compositional Space

Hongshun Chen[a], Ryan Zhou[b], Rujing Zha[a], Zihan Chen[a], Wenpan Li[a], Rowan Rolark[a], John Patrick Reidy[b], Jian Cao[a], Ian McCue[b], Ping Guo[a], David C. Dunand[b], Horacio D. Espinosa[a,*]

[a]Department of Mechanical Engineering, Northwestern University, Evanston, IL 60208.
[b]Department of Materials Science and Engineering, Northwestern University, Evanston, IL 60208.

**Abstract**

High-temperature creep characterization of structural alloys traditionally relies on serial uniaxial tests, which are highly inefficient for exploring the large search space of alloy compositions and for material discovery. Here, we introduce a machine-learning-assisted, high-throughput framework for creep law identification based on a dimple array bulge instrument (DABI) configuration, which enables parallel creep testing of 25 dimples, each fabricated from a different alloy, in a single experiment. Full-field surface displacements of dimples undergoing time-dependent creep-induced bulging under inert gas pressure are measured by 3D digital image correlation. We train a recurrent neural network (RNN) as a surrogate model, mapping creep parameters and loading conditions to the time-dependent deformation response of DABI. Coupling this surrogate with a particle swarm optimization scheme enables rapid and global inverse identification with sparsity regularization of creep parameters from experimental displacement–time histories. In addition, we propose a phenomenological creep law with a time-dependent stress exponent that captures the sigmoidal primary creep observed in wrought INCONEL 625 and extracts its temperature dependence from DABI tests at multiple temperatures. Furthermore, we employ a general creep law combining several conventional forms together with regularized inversion to identify the creep laws for 47 additional Fe-, Ni-, and Co-rich alloys and to automatically select the dominant functional form for each alloy. This workflow combined with DABI experiment provides a quantitative, high-throughput creep characterization platform that is compatible with data mining, composition–property modeling, and nonlinear structural optimization with creep behavior across a large alloy design space.

**Keywords**: Machine learning | high throughput | inverse identification | creep law | alloy

*espinosa@northwestern.edu

**Introduction**

Creep of metals and alloys is a time-dependent plastic deformation under stress that occurs at elevated temperature (usually higher than half of the melting point). Creep mechanisms are typically either diffusion-controlled or dislocation-controlled under different stress levels and temperatures, resulting in permanent plastic deformations (1-5). Central to the development of energy, transport, and aerospace components, the creep behavior of metallic alloys at elevated temperature has been systematically studied for more than one hundred years (1-4, 6-9). Traditional creep experiments involve uniaxial compressive or tensile deformation of a single standard sample under one or more applied stresses and temperatures, which usually last from hours to weeks (10-12). Multiple creep tests under different applied stresses and temperatures are required to obtain the creep properties, which becomes highly inefficient when screening large sets of alloys with different compositions.

To address this challenge, various high-throughput creep experimental techniques have been developed to accelerate the material property characterization. For example, Xu et al. (13) conducted creep tests on a sample with an arc-section exhibiting stress gradients along the tensile direction. The configuration significantly reduced the number of tests required to obtain stress-dependent data. In another study, a cantilever beam bending configuration was adopted to obtain the creep behavior for various stresses at different locations of the beam (14-16). This again leveraged stress gradients within the sample during creep to reduce the required number of tests at constant stresses. Recently, Snitzer et al. (17) fabricated a single dog-bone sample of 316H stainless steel using different additive manufacturing parameters along the longitudinal direction, which allows the simultaneous creep testing of alloys with different microstructures. These high-throughput creep property screenings generated datasets later used to correlate alloy compositions, microstructures, and processing conditions, with creep properties including creep law and creep life. The goal was to achieve fast material discovery and structural design pertaining for high-temperature creep behavior (18-21).

In another investigation, the single plate bulge test was adopted to identify creep properties of alloys while significantly reducing the required number of tests for a given temperature (9, 22, 23). In this configuration, the doming deformation of a thin circular plate, subjected to constant gas pressure on one side, was approximated as a spherical shape to enable fast calculation of the hoop stress and hence determination of creep material properties (22). Using optimization, the creep properties were identified based solely on the bulge apex displacement history (9). Notably, the study suggests that different stress levels within the plate appear sufficient to enable the identification of creep properties using one bulge test at a given temperature (24). As an extension of the plate bulge test, a Dimple Array Bulge Instrument (or DABI) was recently developed (25), to enable high-throughput characterization of alloy creep properties as a function of compositions and microstructure. In this configuration (**Fig. 1**), gas pressure is applied to the back surface of the dimple array to induce plate bulging, while measuring the top-surface displacement history of each dimple using a high-resolution 3D Digital Image Correlation (DIC) technique. Temperature map in the dimple array is measured using an infrared (IR) camera. These accurately measured full-field dimple deformation and temperature enable identification of creep properties using Finite Element Analysis (FEA) and optimization methods (26, 27). However, such workflow of Finite Element Model Updating (FEMU) (28) involves very expensive 3D simulations (c.a. 3 minutes per forward simulation) requiring an excessively long time to identify creep properties.

In this work, we propose the use of a recurrent neural network (RNN) (29), and demonstrate its ability to model time-dependent creep behavior, while dramatically accelerating creep property identification. Previous work has demonstrated the capability of RNN as a data-driven surrogate model of plasticity to capture the yield surface and hardening behavior of composite microstructure (30) but its application to creep law identification is novel. Furthermore, we show that a workflow using RNN is not only highly efficient but also allows flexibility in creep law functional form identification, including one combining several mechanistically driven functional forms previously reported in the literature. Indeed, the formulation here reported enables functional form selection via sparsity regularization with the associated advantage of interpretable physics. By combining high-throughput DABI experiments and a surrogate model, the creep properties of about 50 alloys were identified in less than 2 weeks starting from fabrication, achieving a 100-time acceleration. Such accelerated big data analysis opens new opportunities for material discovery and structural optimization involving creep behavior.

**Inverse identification workflow**

The high-throughput DABI experimental configuration for creep is shown in **Fig. 1a,b, Fig. S1**. Different dimples were fabricated from alloys of various compositions (**Fig. 1a**), which enabled testing creep behavior on alloys with various compositions in parallel. Pressure was applied to the back surface of the dimples to induce doming/bulging deformation at the top surface (**Fig. 1b**). The 3D displacement field of the top surface, captured by means of a 3D DIC setup, was provided to a RNN surrogate model, trained with a large dataset of FEA results, to perform inverse creep property identification (**Fig. 1b,c**).

In this work, we demonstrated the inverse creep property extraction based on Wrought Inconel 625 (Wr-IN625) as baseline, and 47 other alloys with different compositions and unknown creep behavior. We designed a five-step workflow for inverse creep identification of alloys based on DABI experimental data:

(1) formulation of creep laws that can capture the primary creep stage (case 1) and be reduced to standalone conventional creep laws via sparsity regularization (case 2); (2) design of experiments to sample effectively the hyperdimensional parameter space; (3) perform computational modeling and data preparation to create an input-output database for RNN training and validation; (4) perform machine learning to obtain surrogate model analysis of the DABI measurements correlating creep parameters inputs with bulge deformation outputs; (5) inverse identification of creep law based on trained RNN model and experimental data with sparsity regularization.

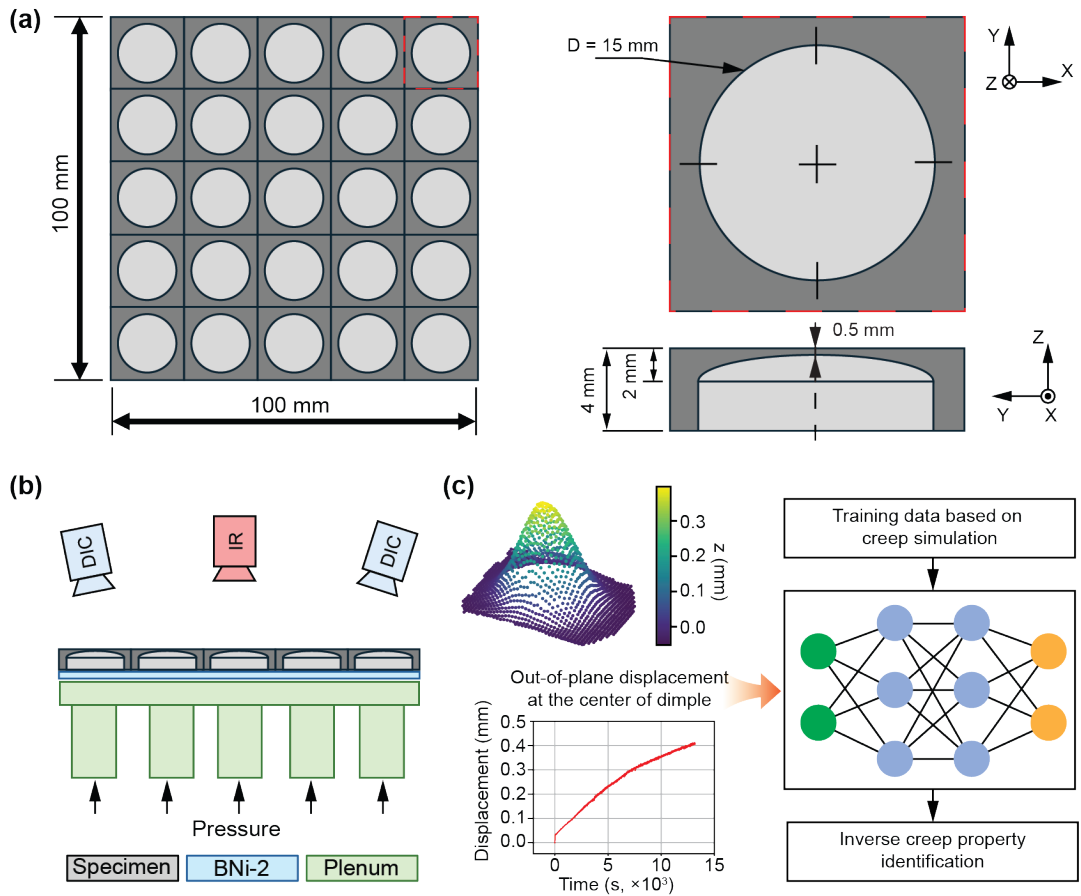

**Figure 1. Schematic of the experimental setup and workflow of creep identification.** (a) Dimple geometry in the dimple array bulge instrumentation with 25 dimples tested in parallel. A solid block was machined to have dimples with an ellipsoidal edge to reduce stress concentration when pressurized. (b) Schematic of the experimental setup of DABI test with DIC and infrared (IR) cameras capturing the displacement and temperature field, respectively. (BNi-2: Nickel-based brazing alloy for diffusion-bonding the sample). (c) Representative 3D bulging deformation and displacement-time profile at the center of the dimple induced by gas pressure. (d) Schematic of pre-trained recurrent neural network, surrogate model, used to inversely identify creep properties.

Two creep laws were proposed for different cases: (case 1) time-dependent creep power law defined in Eq. (1); and (case 2) combined creep law as defined in Eq. (2). The case 1 creep law is given by,

$$\dot{\varepsilon}_{cr} = A\sigma_e^{n(t)}\varepsilon_{cr}^m, \qquad (1)$$

where $\sigma_e$ is the effective stress in the form of von-Mises stress in MPa, $\varepsilon_{cr}$ is the creep strain, $\dot{\varepsilon}_{cr}$ is the creep strain rate in $s^{-1}$, $t$ is time in seconds (s), coefficient $A \in \{0\} \cup [10^{-40}, 10^{-5}]$ with unit depending on the stress exponent $n$ as $MPa^{-1/n}$, and $n(t) = n_\infty + (C + n_0 - n_\infty)e^{-t/\tau_d} - Ce^{-t/\tau_r}$ is the time dependent stress exponent to account for the primary creep behavior as discussed later, $n_0 \in [1,8]$ and $n_\infty \in [1,8]$ can be considered as initial and saturated (i.e., secondary) creep stress exponents while constants $C \in [0,8]$, $\tau_d \in [1000,15000]\,s$, and $\tau_r \in [1000,15000]\,s$ control the transient behavior of the stress exponent as discussed in **Fig. S2**. We also included $\varepsilon_{cr}^m$, the strain dependency with strain exponent of $m$, to expand out model capability. $m = 0$ was used for inverse identification based on experimental data, which reduces the model complexity. This creep law with $n(t)$ is introduced to capture the sigmoidal primary creep behavior found in Wr-IN625 (**Fig. S3**).

The combined creep law for case 2 is defined as:

$$\dot{\varepsilon}_{cr} = A_1 \sinh^{n_1}(\sigma_e/B) + A_2\sigma_e^{n_2}\varepsilon_{cr}^{m_1} + A_3\sigma_e^{n_3}t^{m_2}, \qquad (2)$$

where $A_1, A_2, A_3, B, n_1, n_2, n_3, m_1, m_2$ are the creep parameters. The combined creep law can be reduced to various conventional creep laws as summarized in **Table 1**. For instance, the Norton power law, a classical and general creep law proposed a century ago to capture the power-law type strain-time behavior, can be obtained by appropriate choice of parameters (1). Likewise, the strain hardening power-law dependency on $\dot{\varepsilon}_{cr}$ and $\sigma_e$ originated from diffusion-controlled creep at low stress and dislocation-controlled creep at high stress (5). When creep is diffusion-controlled, the stress exponent $n$ is 1, while $3 \leq n \leq 7$ is typical of dislocation-controlled creep, which occurs under intermediate stress level (31). This power law for creep has been extended to capture the primary creep via the introduction of time exponent ($t^{m_2}$) and strain exponent ($\varepsilon_{cr}^{m_1}$) as shown in **Table 1**. At high stress level, however, the power-law breaks down ($n > 7$), and hyperbolic-sine law (**Table 1**) is usually adopted, which can be attributed to significant dislocation climb (4, 5).

**Table 1**. Standalone conventional creep laws that can be recovered by setting some parameters equal to 0 in Eq. (2). The parameter bounds used during creep law identification are also provided.

| Creep law | Functional form for $\dot{\varepsilon}_{cr}$ | Bounds |
| --- | --- | --- |
| Norton power law (PL) | $A_2\sigma_e^{n_2}$ | $A_2 \in \{0\} \cup [10^{-40}, 10^{-5}]$ |
| | | $n_2 \in \{0\} \cup [1,8]$ |
| Strain-hardening power law (SPL) | $A_2 \sigma_e^{n_2} \varepsilon_{cr}^{m_1}$ | $m_1 \in [-1,0]$ |
| Norton-Bailey time-hardening power law (TPL) | $A_3 \sigma_e^{n_3} t^{m_2}$ | $A_3 \in [10^{-40}, 10^{-5}]$<br>$n_3 \in \{0\} \cup [1,8]$<br>$m_2 \in [-1,0]$ |
| Hyperbolic-sine law (HS) | $A_1 \sinh^{n_1}(\sigma_e/B)$ | $A_1 \in [10^{-40}, 10^{-5}]$;<br>$n_1 \in \{0\} \cup [1,8]$<br>$B \in [10^0, 10^3]$ |

Together with elastic modulus ($E \in [50,300]\ GPa$), 7 and 10 parameters are associated with the material properties for case 1 and case 2, respectively. With 3 more input parameters associated with the experimental configuration (i.e., $T_j \in [2,10]\ hour\ (h)$, $P_1 \in [3,25]\ MPa$, $P_2 \in [3,25]\ MPa$, $P_2 \geq P_1$), the input space has 11 (case 1) and 13 (case 2) parameters, respectively. For both cases, we used a Sobol sequence to uniformly generate samples based on the specified bounds. During generation, the samples are subject to selection based on the criteria of strain rate and total creep strain to avoid significantly large or small creep strain (rate) under small or high stress, respectively, while the generated data for training was also selected again to remove those having no creep deformation (See selection criteria in **Materials and Methods**). The final samples for each case in the RNN training and validation set are shown in **Fig. S4-5**. After selection, 45,844 and 155,186 valid samples were used in case 1 and case 2, respectively.

To generate output data, ABAQUS 2022 (Dassault, France) was used with Sobol selected input parameters. The creep laws in Eq. (1) and (2) were implemented in ABAQUS using the CREEP Fortran subroutine. The simulation model is a reduced-order axisymmetric model to predict displacement outputs (**Fig. 2a**). Instead of directly using the displacement history at all locations of the dimple top surface, as the output for model training, the displacements were fitted with a Generalized Gaussian function further reducing the output dimensions, as in Eq. (3) (**Fig. 2b**):

$$D(x,t) = D_{amp}(t) \cdot exp\left(-\left(\frac{x}{S_d(t)}\right)^{S_p(t)}\right), \tag{3}$$

where $D(x,t)$ is the displacement (mm) at location $x$ and time $t$, $D_{amp}(t)$ is the amplitude at time $t$ which is equivalent to the displacement at the dimple apex, $S_d(t)$ and $S_p(t)$ are the standard deviation and the shape parameter, respectively, at creep time $t$. Therefore, $D_{amp}(t), S_d(t), S_p(t)$ are the outputs as a function of time, which are later used during machine learning model training.

We used a pressure-jump experimental profile with different pressure levels at different time intervals to deform the dimples, mirroring the conventional stress-jump creep test (**Fig. 2c**) (32). The pressure-jump profile is defined by two pressure levels as $P_1$, and $P_2$ ($P_2 \geq P_1$) and a parameter $T_j$ corresponding to the time at which the pressure jumps. This pressure-jump profile can maintain or enrich the stress levels throughout the thickness of the dimple, providing sufficient information of top-surface displacement history for robust inverse creep property identification (**Fig. 2c**).

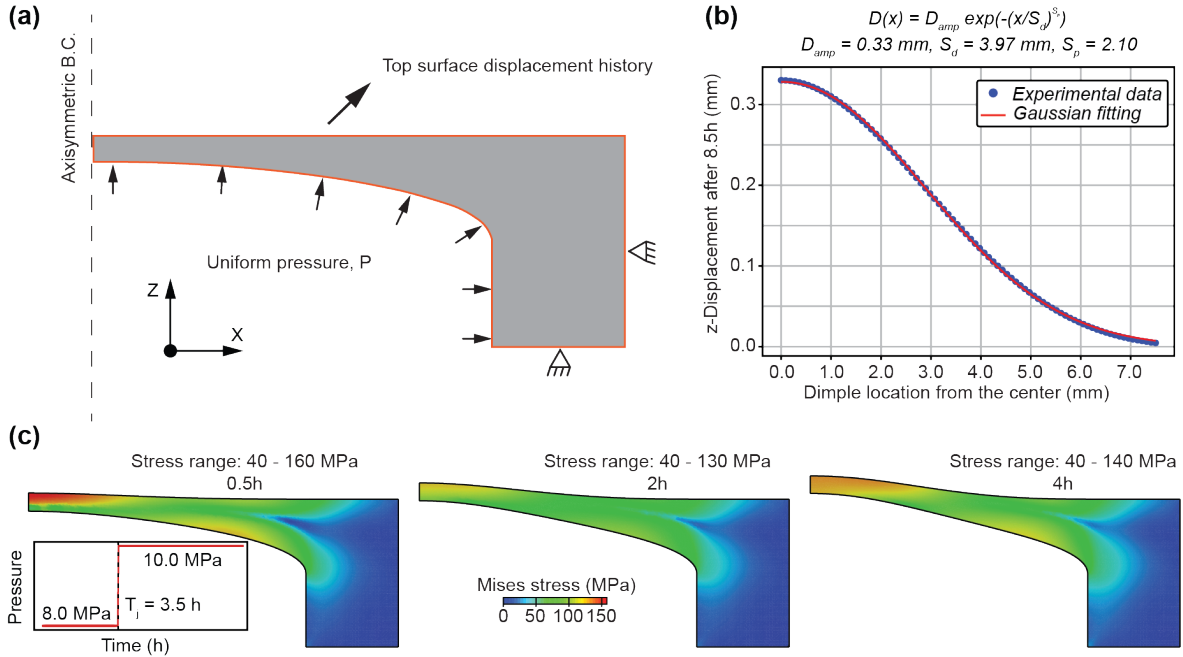

**Figure 2. Simulation model setup for data generation.** (a) Dimple geometry with axisymmetric boundary condition. (b) A representative simulation result of the bulge deformation after 8.5 h of creep fitted with a generalized Gaussian equation. (c) Stress distribution within the dimple under pressure jump (8 to 10 MPa) test showing a range of von-Mises stress from 40 to 160 MPa for three times (i.e., 0.5, 2 and 4 h). This example case has the following parameters for Eq. (1): $E = 150\ GPa$, $A = 7.4 \times 10^{-17}$, $C = 2.7$, $n_0 = 4.9$, $n_\infty = 4.3$, $\tau_d = 13823.6\ s$, $\tau_r = 8090.4\ s$).

For simplicity and for overfitting mitigation, the RNN model used in this work has gated recurrent units (GRU, **Fig. S6**), which have a simpler formulation and fewer trainable parameters compared to Long Short-Term Memory (LSTM) (33). The schematic of the surrogate model (**Fig. 3a**) shows that the GRU takes temporal and non-temporal input parameters and the Gaussian-fit outputs for training. The pressure $P_1$ changes to $P_2$ at $T_j$, while other nontemporal input parameters (i.e., the material parameters) remain unchanged in all recurrent time series (**Fig. 3a**). Before feeding the training, a $log_{10}(\cdot)$ operator was applied on parameters of $A, A_1, A_2, A_3, B$ to ensure a similar order of magnitude among input parameters' values. Subsequently, all inputs and outputs were standardized with their respective means and standard deviations. The samples were split into training and validation (unseen) datasets with a split ratio of 8:2. The loss function ($\mathcal{L}$) employed during RNN training is given by the relative squared error (RSE) function defined by,

$$\mathcal{L} = \frac{\sum_i^t (O_{pr}^i - O_{tr}^i)^2}{\sum (O_{tr}^i - \overline{O}_{tr}^i)^2}, \tag{4}$$

where $O_{pr}^i$ is the predicted output vectors of $D_{amp}, S_d, S_p$ at the $i$-th time step, $O_{tr}^i$ is the true output vectors at the $i$-th time step, and $\overline{O}_{tr}^i$ is the averaged output vectors at the $i$-th time step. Despite training with $\mathcal{L}$ as a single metric, we also quantified other metrics to measure the performance of trained model as discussed later.

The inverse identification of creep laws was conducted using the gradient-free Particle Swarm Optimization (PSO), which enables the efficient global search of the input space for global minima (34). The PSO algorithm shown in **Fig. 3b** navigates the input space, utilizes the trained RNN model to evaluate the loss of candidates, and finds the solution that yields minimal loss. Details of the PSO algorithm are provided in **Table S1**. For case 1, we will also demonstrate the identification of temperature-dependent behavior of the functional form given in Eq. (5), which requires no additional training data (4).

$$\dot{\varepsilon}_{cr} = A\sigma_e^{n(t)} \exp\left(-\frac{Q}{RT}\right), \tag{5}$$

where $Q$ is the activation energy in $J \cdot mol$, $R$ is the gas constant of $8.31 \frac{J}{K \cdot mol}$, and $T$ is the temperature in Kelvin (K). In addition, as the combined law (Eq. (2)) is a combination of different functional forms of creep laws, $L_2$ sparsity regularization is introduced during creep identification compared to those identified without regularization (**Materials and Methods**).

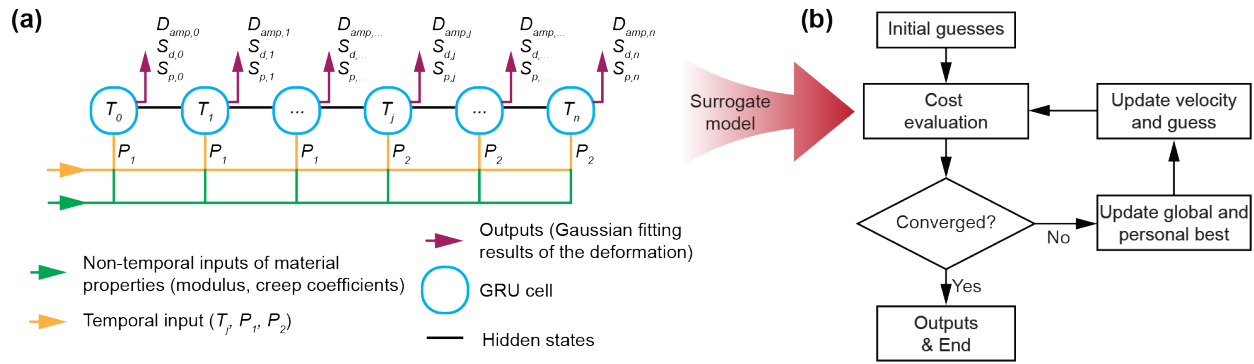

**Figure 3. Schematic workflow of surrogate model and inverse property identification method using particle swarm optimization.** (a) The architecture and input-output pairs of the RNN model. (b) The workflow of the inverse creep property identification using a gradient-free global optimizer and the trained RNN surrogate model.

**Results**

*Case 1: Creep law with time-dependent stress exponent*

The creep law for case 1 is defined in Eq. (1) with time-dependent stress exponent to capture the primary creep behavior of Wr-IN625. The input-output dataset constructed by simulation data was fed to train the RNN, achieving the lowest loss of $\mathcal{L} = 0.0019$ for the training dataset and $\mathcal{L} = 0.0039$ for the validation dataset using a hidden size of 128 and 3 layers (**Fig. 2c, Fig. S7**). A hyperparameter sweep was conducted to investigate the effect of the number of hidden layer, hidden size, sample size, and cell type (i.e., LSTM or GRU) on training outcome (**Fig. S7**). We also evaluated the trained model using mean squared error (MSE) and mean absolute error (MAE), which further confirmed strong training performance (**Fig. S8**). **Fig. 4** shows the predicted outputs compared to the true outputs of unseen data in the validation dataset at 3 selected time steps (i.e., 7,500, 22,500, and 37,500 s) where the linearly fitted slope demonstrates an overall accuracy of > 99%. For very rare cases of smaller shape parameter $S_p$ (c.a. 0.001% of training samples), indicating a sharper shape of the Gaussian function, the predictions are less satisfactory (**Fig. 4**).

To illustrate the capabilities of the surrogate model, we report predicted and identified creep behaviors. **Fig. 5a** shows predicted and true outputs of two representative cases in unseen datasets with different curve behaviors and different amplitude at the end of creep, further demonstrating the good training performance. For the case in **Fig. 5a-i**, the displacement remains nearly flat initially due to the high value of $\tau_r = 10100\ s$ (slow rise time) and low $\tau_d = 1280\ s$ (fast decay time) (**Table S2**). The predictions of outputs for this case have small spikes in the first few steps (< 1% difference) and then stabilize, providing good agreement. The large pressure jump introduces another spike at $T_j = 6\ h$, which is also well-captured by the surrogate model with a relative difference less than 3% for $D_{amp}(T_j)$. For the case in **Fig. 5a-ii**, the prediction becomes smooth despite the pressure jump, indicating that the RNN can capture the smooth response with high accuracy.

Next, we evaluated inverse identification performance on unseen validation data, using the PSO algorithm outlined in **Fig. 3b**. We found that the inversely identified inputs provide good agreement of predictive outputs and targeted outputs ($\mathcal{L} \sim 10^{-3}$, **Fig. 5b**). The inversely identified input parameters have some discrepancy compared to the true input parameters for combined creep law (Eq. (2)), suggesting the non-uniqueness of the solution (**Table S3**). More about predictive and inverse identification performances, based on the unseen datasets, are reported in **Fig. S9-10, Table S4,** respectively.

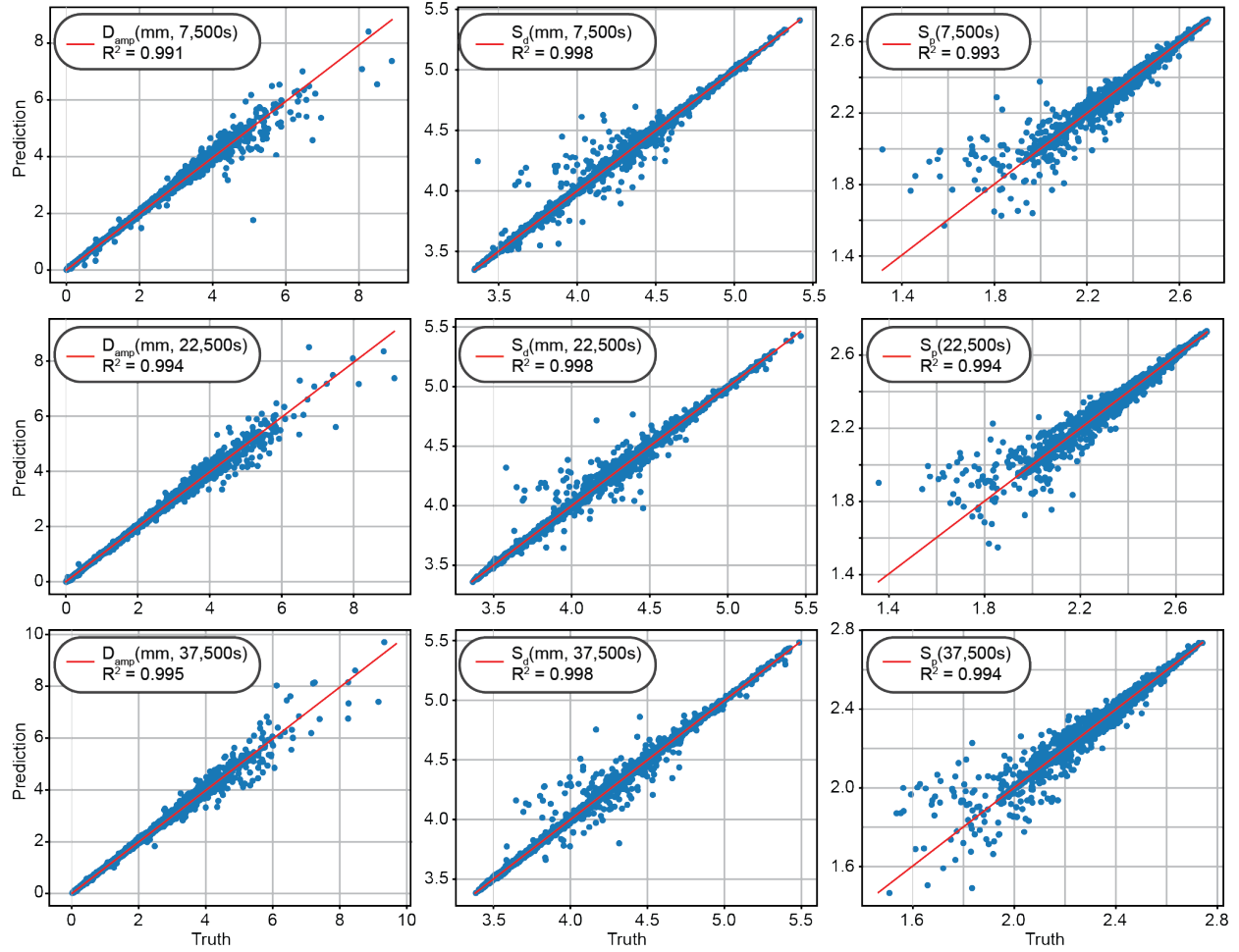

**Figure 4. Performance of the trained RNN as the surrogate model in predicting unseen creep properties and pressure profiles.** Comparison between true and predicted amplitude ($D_{amp}$), standard deviation ($S_d$), and shape parameters ($S_p$) at different time steps (i.e., 7,500, 22,500, 37,500s, respectively).

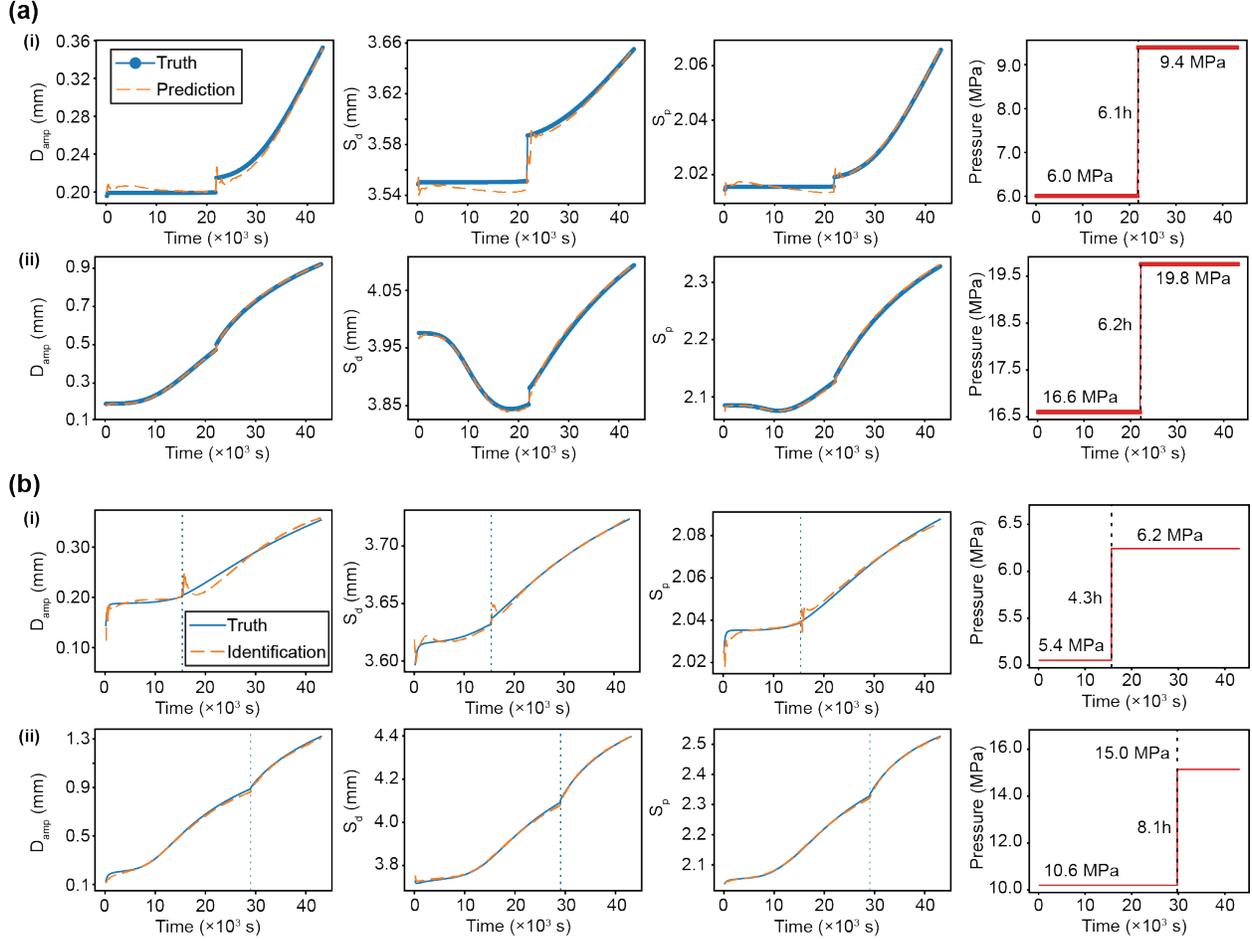

**Figure 5. Representative prediction and inverse identification of the output histories.** (a) Comparison between true and predicted history of fitted amplitude ($D_{amp}$), standard deviation ($S_d$), and shape parameters ($S_p$) and the corresponding pressure profiles. (b) Comparison between true and predicted amplitude ($D_{amp}$), standard deviation ($S_d$), and shape parameters ($S_p$) based on the inversely identified creep parameters.

Leveraging the trained surrogate model, we conducted inverse identifications based on DABI experimental data on Wr-IN625. Data preprocessing on the acquired 3D DIC data generated the 2D Generalized Gaussian fitting results at different time steps (**Fig. 6a**, **Materials and Methods**). The 2D visualization of the fitting results along with the data points after rotating to the quarter cross section shows a good agreement between them (**Fig. 6b**). Since DABI creep tests were performed on Wr-IN625 under different temperatures, we identified parameters in the creep law with time-dependent stress exponent (Eq. (1)) along with a temperature-dependent term of Arrhenius functional form for the activation energy as in Eq. (5). The processed outputs and known $P_1 = 8\ MPa$, $P_2 = 8\ MPa$, and $T_j = 2\ h$ were fed to the inverse identification workflow (**Fig. 3**). Here, since $P_1 = P_2$, the selection of $T_j$ is arbitrary. To improve accuracy, we first identified the temperature-dependent elastic modulus, following a Wachtman's equation $E_T = 177.5 - 0.21T \exp\left(-\frac{948.5}{T}\right)$, where $E_T$ is the elastic modulus (GPa) at temperature $T$ (K), based on the initial 15 timesteps of creep (i.e., first 2250s) as shown in **Fig. 6c** (35, 36). The identified modulus is slightly smaller than the Young's modulus tested using Joule-heated tensile bars made of Wr-IN625 (**Fig. 6c**). Fixing the identified modulus, the identified creep law based on 9 different dimples under different temperatures are shown in Eq. (6) with $\mathcal{L} = 0.111$, indicating a goodness of fit of 88.9% (More details of inverse identifications in **SN5, Fig. S11**).

$$\dot{\varepsilon}_{cr} = 4.2 \times 10^5 \sigma_e^{4.3+1.29e^{-t/7818.4}-1.9e^{-t/4753.9}} \exp\left(-\frac{4.23 \times 10^5}{RT}\right), \qquad (6)$$

where $n_\infty = 4.3$ is considered the stress exponent for steady-state creep. The strain rate vs. stress of identified steady-state creep law (with stress exponent of $n_\infty$) based on DABI experiments is c.a. 2 times smaller that of steady-state creep law obtained via conventional compressive creep experiments, i.e., $\dot{\varepsilon}_{cr} = 1.32 \times 10^6 \sigma_e^{4.8} \exp\left(-\frac{4.61 \times 10^5}{RT}\right)$ at 750, 800, and 850°C (**Fig. 6d**). Possible sources for the discrepancy are sample size effects, grain morphology variations, and temperature differences used in the measurements. While compressive creep cylinders are polycrystalline, Wr-IN625 dimples tested via DABI experiments may not possess a consistent number of grains through the thickness. Furthermore, dynamic recrystallization during loading could further complicate the effects of grain and sample size. Previous work indicates that for specimens under 1 mm in scale (approximately 5 grains or fewer in the sample thickness or width) exhibits a size-weakening effect, similar in magnitude to the creep strain rate differences between the baseline compressive result and DABI experiments (**Fig. 6d**) (37). Despite inherent uncertainties for strain rates obtained via most creep testing methods, measured strain rates remain relatively consistent across DABI-tested Wr-IN625 dimples of similar temperature. Moreover, the predicted displacement profiles at different locations of representative dimples under different temperatures, obtained based on the identified creep law, (i) show good agreement with the experimental data, and (ii) capture the initial primary creep behavior with the rising time of stress exponent $\tau_r = 4753.9\ s$ in good agreement with the observation that the sigmoidal primary creep ends at approximately 5,000 s (**Fig. 6e**). Furthermore, we showed in **Fig. 6f** that the identified creep law in Eq. (6) generalize well with the experimental data not covered by the cases used in inverse identifications such as pressure-jump experiment with $P_2 = 10\ MPa$ at $T_j = 4\ h$ and temperature of 712°C. See more experimental data used in inverse identification and test as well as RNN-predicted outputs (i.e., $D_{amp}, S_d$, and $S_p$) and displacement history based on identified creep law (Eq. (6)) in **Fig. S12-14**. It is noteworthy that enforcing the temperature-dependent functional form in Eq. (6) decreases the identification accuracy as the stress exponent is also temperature-dependent: the higher the temperature, the less the primary region (**Fig. 6e**). Therefore, we also showed that via individual identifications, the fitted creep laws with time-dependent stress exponent achieve excellent agreement between predicted and experimental displacements ($\mathcal{L}$ is in the order of 0.001, **Fig. S15**). We note that these results can later be used in discovering new temperature-dependent functional forms for creep constitutive law.

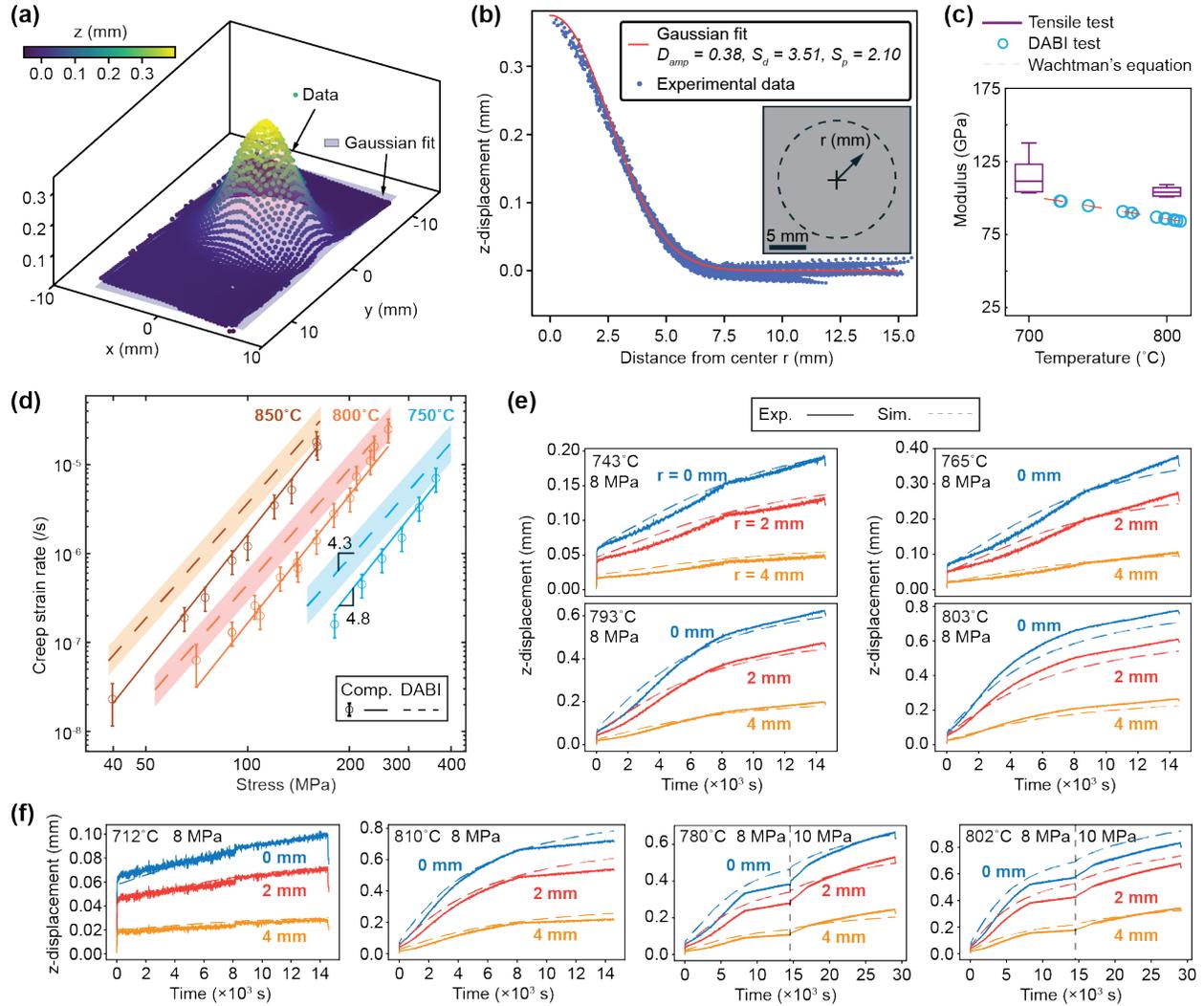

**Figure 6. Temperature-dependent creep property identification based on DABI experiments for Wr-IN625.** (a) Representative comparison between experimental displacement in the z-direction and 3D generalized Gaussian fitting result. (b) A quarterly cross-sectional view of the Gaussian fitting result compared to the experimental data. (c) Identified temperature-dependent modulus following Wachtman's equation. (d) Double logarithmic plot of creep strain rate versus effective stress for the inversely identified creep parameters and creep parameters obtained from conventional compression creep, respectively. The color-shaded transparent error bands were obtained based on ± 10°C temperature variation recorded in DABI tests. The discrepancies with the compressive creep data are estimated to be 30% if the strain rate is above $10^{-7}s^{-1}$ and 50% if below $10^{-7}s^{-1}$. (e) Comparison between representative experimental displacements (constant pressure of 8 MPa) at different locations of r = 0, 2, 4 mm, respectively, and simulation results at four temperatures based on the identified creep law. (f) Four other representative experimental cases, not used in inverse design but used for testing, including two cases with pressure-jump at $T_j = 4\ h\ or\ 14400\ s$, with $P_1 = 8\ MPa$, and $P_2 = 10\ MPa$. Note that the data with $P_1 = 8\ MPa$, in the last 2 plots, were used in inverse identification, while the $P_2 = 10\ MPa$ were used to test whether the identified creep law generalizes well with unseen data.

*Case 2: Creep properties of 47 different alloys*

In this section, we demonstrate the use of the RNN trained with combined creep law as in Eq. (2) to achieve: (i) high-throughput creep law identification of 47 different alloys based on the DABI experiments, and (ii) automatic creep law selection with sparsity regularization for the purpose of interpretability.

We began by training the RNN model, examined the predictive performance on unseen simulation data, and demonstrated the inverse identification capability (**Fig. S16-S20, Table S5**). We confirmed that the following hyperparameters: 5 hidden layers, 512 hidden sizes, and 155,186 samples, provide the best training performance. Without imposing sparsity regularization, the inversely identified combined creep laws based on unseen simulation data are non-unique compared to the ground truth, while the inverse identification based on standalone creep laws (i.e., HS, PL, SPL, TPL in **Table 1**) provided unique solutions (**Fig. S16-S20, Table S5**).

With an experimental dataset of 47 different alloys, we performed inverse identification employing the combined creep law of Eq (2), without and with sparsity regularization (Loss function of $\mathcal{L}'$ with sparsity regularization term as in Eq. S4, **SN6**). **Fig. 7a** shows 3 representative inverse identification results with sparsity regularization of Fe-rich, Ni-rich, and Co-rich alloys, respectively. The compositions are given in **Table S6**. The results demonstrate agreement between predicted and true displacement-time profiles, at different locations of the dimples, for the three alloy compositions. Here, we classified different alloys based on compositional fractions of principal elements, i.e., Fe-rich, Ni-rich, and Co-rich, indicating the dominant element has highest weight fraction (**Table S7**). It is worth noting that, despite the training data having a bound of $T_j \in [2,12]$ h, the inverse identification accurately extrapolates to the cases with end creep time of ~20 minutes (**Fig. S21**). This provides further evidence that the RNN learned from the dataset without overfitting and was capable of inverse identification based on unseen experimental data.

Despite sparsity regularization, the loss value (only the $\mathcal{L}$ part) is comparable to those identified without sparsity regularization (**Fig. 7b**). While sparsity regularization always yielded simplified creep laws, which allow better interpretability, in terms of operating deformation mechanics, and uniqueness of the identified creep laws, which are summarized in **Table S8**.

Analysis of the identified creep laws, for the 47 different alloys, reveals patterns of dominant creep laws (**Fig. 7c**). The HS law predominantly describes the Fe-rich alloys well, while PL and SPL creep law have similar fractions in describing the behavior of Co-rich alloys (**Fig. 7c**). For Ni-rich alloys, no single type of creep law appears to govern their behavior (**Fig. 7c**). Several alloys in each elemental bin may be further classified as multi-principal element alloys with near-equimolar Fe, Co, Ni, or Cr: these alloys may result in mixed behavior and warrant further study. Overall, these results demonstrate that when machine-learning techniques are combined with high-throughput creep experiments, the data mining and hence discovery of creep behaviors for a multitude of alloy composition becomes possible.

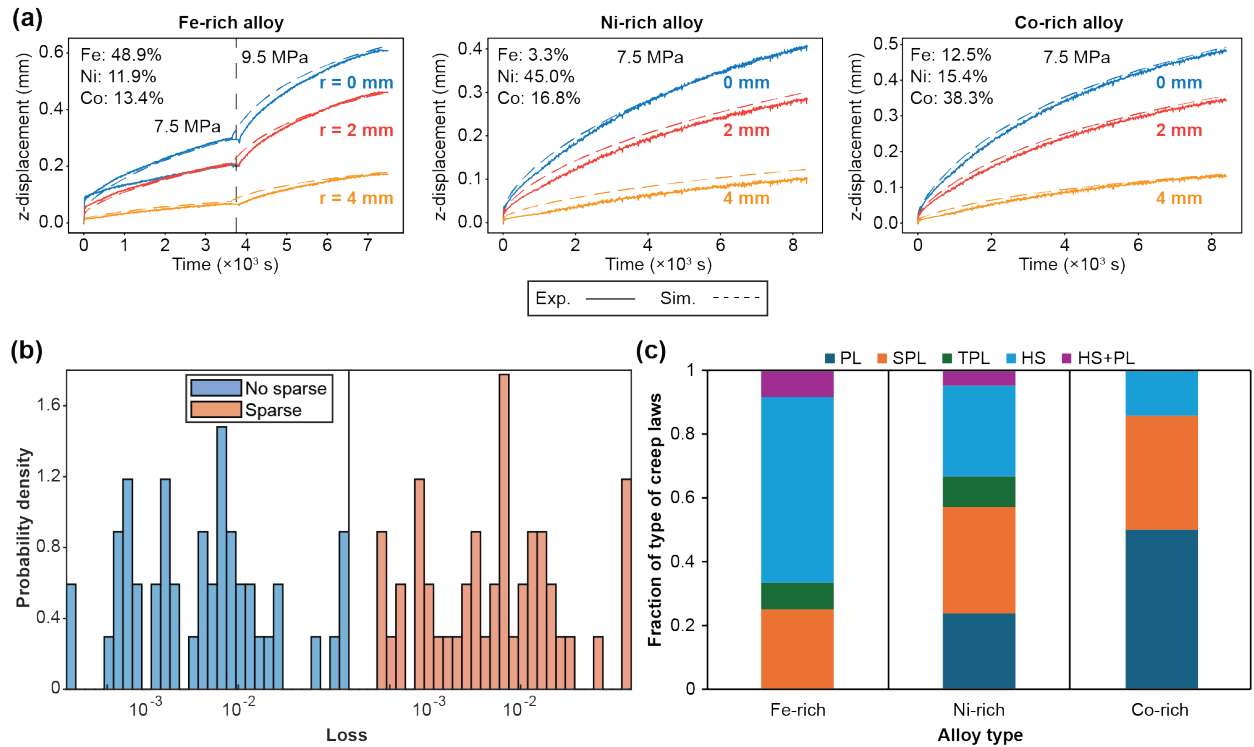

**Figure 7. Summary and analysis of creep law identification of 50 unknown creep laws with sparsity regularization.** (a) Comparison between experimental and predicted displacement histories at r = 0, 2, 4 mm, for selected examples of Fe-rich, Ni-rich, and Co-rich alloys, respectively (elemental fractions given in wt. %, full composition in Table S7). Their respective creep laws are: Fe-rich: $\dot{\varepsilon}_{ce} =$

$3.16 \times 10^{-7} \sinh^{4.15}\left(\frac{\sigma_e}{87.16}\right)$; Ni-rich: $\dot{\varepsilon}_{ce} = 5.5 \times 10^{-14} \sigma_e^{3.28} \varepsilon_{cr}^{-0.32}$; Co-rich: $\dot{\varepsilon}_{ce} = 1.2 \times 10^{-16} \sigma_e^{5.2}$. (b) Comparison of loss between identifications with and without sparsity regularization. (c) Dominant identified creep law with sparsity regularization for the three different classes of alloys.

**Discussions**

Improving efficiency and accuracy of property identification processes based on novel experimental techniques and machine learning approaches has long been of interest to experimentalists aiming to accelerate materials discovery and structural optimization. In the present work, we leveraged RNN as a surrogate model to accurately correlate the time-dependent creep deformation of pressurized dimples made of metal alloys. When these surrogate models are used in combination with high-throughput DABI experiments, the approach enables parallel creep testing of alloys with various compositions and efficient identification of creep laws with high accuracy.

Instead of using a conventional finite element optimization framework to iteratively update material parameters, our data-driven machine learning model bypasses expensive finite element simulations in the loop and thus significantly accelerates the inverse identification process (~5 min per dimple on an NVIDIA RTX 5000 Ada with 300 initial particles and 100 PSO iterations). Moreover, the gradient-free PSO optimizer exploits the efficient surrogate RNN model (forward evaluation time of each candidate solution and each recurrence is ~$10^{-4}$ s) to compute the loss and enables a global search in the high-dimensional parameter space. The high training accuracy of the RNN indicates that the surrogate model captures the time-dependent creep response well, while the low dimensionality of input parameters due to the combinatorial use of conventional creep laws significantly alleviates the ill-posedness of the inverse problem. In addition, the non-uniqueness of the creep-law solution is further mitigated by sparsity regularization: inverse identification based on standalone HS, PL, SPL, and TPL creep laws recovers the corresponding true dominant solutions (**Table S6-7**). Therefore, by using the combined creep law in Eq. (2) together with sparsity regularization, we improved on both interpretability and uniqueness of the identified solution. Furthermore, the creep laws defined in Eqs. (1) and (2) naturally allow the extraction of temperature dependence in a wide range of functional forms, rather than being restricted to the classical Arrhenius form, without requiring additional training data in an enlarged input space. As a result, the surrogate RNN model also supports symbolic regression of temperature dependence to discover new closed-form phenomenological creep laws. One example is shown in **Fig. 6e** for Wr-IN625, where the sigmoidal primary creep behavior and the time-dependent stress exponent clearly depend on temperature: at lower temperatures (700–750°C), the sigmoidal primary creep is more pronounced, whereas at higher temperatures (750-810°C) the response approaches a more classical monotonic decay in creep rate.

Finally, beyond the specific combined law in Eq. (2), the additive formulation can be extended to include additional conventional creep laws, providing an efficient, automatic law-selection strategy that identifies suitable creep models for different alloys in a large creep dataset. This is also consistent with the guidelines of Holdsworth *et al.*, which explicitly link creep law selection to material characteristics, data distribution, and intended application (i.e., models with a sound physical basis for materials scientists vs. effective modeling for design engineers) (8).

We proposed the creep law given in Eq. (1), to capture the primary creep behavior of Wr-IN625 via a time-dependent stress exponent (**Fig. 6e**). In such law, the initial low creep rate is followed by a transient acceleration and subsequent deceleration, i.e., a sigmoidal primary creep response consistent with prior observations in additively manufactured IN625 at 700–800 °C (32, 38). Because the alloy was homogenized prior to testing, additional dynamical recrystallization during creep is unlikely. Overall testing times for both DABI and baseline compressive creep experiments were kept as short as practical to limit additional microstructural evolution, including γ″ precipitation and subsequent transformation to δ phase (39). However, the complex multiaxial stress state imposed by the DABI geometry (**Fig. 2c**) may promote localized abnormal grain growth in the most highly stressed regions, leading to an oligocrystalline microstructure within the deformed dimples. Concurrent grain coarsening and locally accelerated primary creep in these regions are therefore a plausible origin of sigmoidal primary creep behavior. In our DABI data, this sigmoidal character is most pronounced at lower temperatures and becomes progressively weaker as temperature increases, with the primary stage approaching a more conventional monotonic decay in creep rate at the highest test temperature (**Fig. 6e**), suggesting that the phenomenon is confined to an intermediate stress–temperature window where dislocation accumulation and recovery occur on comparable time scales. Notably, a clear sigmoidal primary creep effect is not observed in the other 47 tested alloys. A likely explanation is that the initial equiaxed grain structure and stored dislocation density in the wrought plate differ substantially from those in the direct energy deposition (DED) processed specimens: even after homogenization at 1150 °C, residual dislocation networks generated by prior rolling and dimple machining may persist in Wr-IN625, whereas the DED alloys start from a columnar microstructure, a differing distribution of secondary phases and dislocation density, and lower stored energy. In addition, Wr-IN625 is particularly susceptible to sigmoidal primary creep at our testing temperatures because its high Mo+Nb content enhances solid-solution strengthening and planar slip, facilitating rapid local dislocation accumulation upon rolling and the formation of a cellular substructure (38). Further detailed microstructural characterization before and after DABI testing (e.g., grain size, subgrain/cell structure, dislocation density, and precipitate state) will be required to fully ascertain the operative primary creep mechanisms in Wr-IN625. Such characterization is left to future work.

Each dimple in the DABI configuration, when subjected to back-surface pressure, develops rich, multiaxial stress states and strong through-thickness stress gradients. When combined with precise temperature control and in situ temperature measurement, this configuration provides a robust foundation for inverse identification of creep properties using top-surface deformation fields measured via 3D digital image correlation (3D-DIC). Here, we demonstrate these capabilities by identifying creep laws using Wr-IN625 as a baseline material. We further show that test parallelization across multiple materials, together with the availability of rich stress information within each dimple, dramatically reduces experimental time relative to conventional uniaxial creep testing, which requires separate experiments at multiple stress levels. Specifically, by testing 25 materials simultaneously under three effective stress levels within the dimples, an acceleration factor of at least 25 × 3 = 75 is achieved compared to traditional approaches. In this context, the combination of accelerated creep experimentation and machine-learning-based creep law identification enables systematic data mining of creep properties across large alloy spaces with varying compositions. In the present study, we identified creep properties for 47 distinct alloys over a range of temperatures using sparsity-regularized learning, which automatically selects the most appropriate creep law to fit the experimental data and accurately capture the observed creep behavior (Fig. 7c).

While we classified alloys by the dominant elemental constituent in the present work, this represents only one of several possible descriptors. We anticipate that, as the database expands, the relationship between alloy composition and creep response can be further refined using interpretable neural networks and symbolic regression techniques to derive closed-form constitutive expressions (40-44). The learned correlation between alloy composition and creep response is analytically differentiable, enabling efficient gradient-based structural optimization with temperature-dependent and spatially graded material properties. This capability can significantly accelerate design and performance optimization across a wide range of applications, including aerospace propulsion and hypersonic systems (turbines, combustors, rocket engines, hypersonic skins), structural components for nuclear energy (fission, fusion, and small modular reactors), biomedical implants and bio-interfaces, and defense and impact-resistant structures, among others.

**Conclusions**

The present work establishes a pathway for high-throughput creep characterization and model discovery by combining dimple array bulge instrument experiments, a recurrent neural network surrogate with Gated Recurrent Unit, and global optimization. The proposed creep law with a time-dependent stress exponent rationalizes the sigmoidal primary creep behavior of wrought IN625, while the creep law combining various conventional creep models, coupled with sparsity regularization, enables automatic selection of dominant constitutive forms across 47 additional Fe-, Ni-, and Co-rich alloys. By replacing traditional finite element optimization loops with a fast, data-driven surrogate, the framework reduces the time required for inverse identification to minutes per dimple and accurately extrapolates beyond the training space. The resulting creep laws show good agreement with conventional uniaxial creep data and, more importantly, provide a scalable pathway for building composition–creep property databases suitable for data mining and interpretable symbolic regression. We anticipate that extending this approach to broader alloy classes and more complex loading histories will enable creep-informed structural optimization and accelerate the discovery of high-temperature alloys for energy, biomedical, and aerospace applications.

**Materials and Methods**

**Data generation and preparation**

A database with input-output pairs of creep properties and responses was built for training the RNN model. First, a Sobol sequence was used to generate numerous samples of inputs with 11 or 13 dimensions, depending on the types of creep laws (Eq. (1) and (2)), including the experimental configuration parameters of $T_j$, $P_1$, and $P_2$ (45). During sampling, to avoid significantly large or small strain, the following criteria were used to determine whether the sample was discarded for both cases, Eqs. (1) and (2), respectively:

- *Criteria 1*: if $\dot{\varepsilon}_{cr} < 10^{-20} s^{-1}$ when substituting $\sigma_{eff} = 600\ MPa$, $\varepsilon_{cr} = 0.01$, $t = 10000\ s$ into the creep strain rate equation, the sample is discarded.
- *Criteria 2*: if $\dot{\varepsilon}_{cr} > 1 s^{-1}$ when substituting $\sigma_{eff} = 50\ MPa$, $\varepsilon_{cr} = 0.01$, $t = 10000 s$ into the creep strain rate equation, the sample is discarded.

After simulation to obtain the output, the samples were selected again based on the following criteria:

- *Criteria 3*: the creep simulation needs to last at least 1 h and converge.
- *Criteria 4*: the total apex displacement (i.e., fitted $D_{amp}$) after initial pressure ($P_1$) applied until $T_j$ and second pressure ($P_2$) applied until the end of simulation ($T_{end} = 12\ h$) needs to be higher than 0.01 mm.
- *Criteria 5*: the fitted amplitude at $T_{end}$ is higher than 0.01 mm, approximately twice the resolution of DIC in the out-of-plane direction.

The samples were input into the ABAQUS subroutine developed for the combined creep law in Eq. (1) and (2) to generate the output database. A MATLAB script was developed to automatically interact with ABAQUS (including the generation of input file and subroutine file, and job submission) and run batch simulations based on the availability of CPU processors and sampled inputs. Each simulation took 1-3 min using Intel(R) Xeon(R) Gold 6338 CPU @ 2.0 GHz. The dimple dimensions are shown in **Fig. 1a**, and the simulation model is shown in **Fig. 2a**. The axisymmetric simulation model has 2117 CAX4R elements. The total simulation step time ($T_{end}$) was set to 12 h, which is longer than $T_j$, limited by the bounds (**Table 1**). The interval of output was set to 150 s.

A 3D Digital Image Correlation (DIC) setup with two cameras was used to capture the bulging displacement of the dimples. This was used as ground truth for inverse creep property identification (**Fig. 1b**). The VIC-3D software was used to perform DIC analysis based on left and right images (**Fig. 1b**). The displacement data based on DIC analysis was interpolated to have interval of 150 s and subsequently fitted using 2D Generalized Gaussian equation given by

$$D(x,y,t) = D_{amp}(t) \cdot exp\left(-\left(\frac{x-x_c}{S_d(t)}\right)^{S_p(t)} - \left(\frac{y-y_c}{S_d(t)}\right)^{S_p(t)}\right), \tag{7}$$

where $x_c$ and $y_c$ define the center location of the fitted Gaussian. Example fitting results are given in **Fig. 6a,b**.

**RNN model: Training and inverse identification**

The RNN model shown in **Fig. 3a** was developed using the Pytorch library as a surrogate to replace the expensive simulations. The total epoch of training is 500 based on various hyperparameters (**Fig. S7 and S14**) (46). The gradient-based optimizer Adam was used to update the trainable parameters of GRU (**Fig. S5**) (47). The loss of validation set was monitored during the training process, which informs the scheduler (ReduceLROnPlateau) with a patience of 15 epoch. The initial learning rate was 1e-3, which was reduced by the scheduler if loss did not improve in number of epochs defined by patience. The step-like decrease in loss is a result of the reduction of learning rate determined by the scheduler (**Fig. S7 and S16**).

The global optimizer PSO, as shown in **Fig. 3b,** was used during the inverse identification of creep properties, leveraging the trained RNN surrogate model to evaluate loss (i.e., the error between ground truth and predicted deformation). During inverse identification based on simulation data, the loss is the same as $\mathcal{L}$ in Eq. (4). During inverse identification based on experimental data, the loss function was modified to introduce weights to different outputs (Eq. (S3)). High weight of 0.99 was given to the loss contributed from the amplitude ($D_{amp}$) as shown in Eq. (S3) due to the dominantly high values of loss of $S_d$ and $S_p$. Despite weighing significantly to $D_{amp}$, we checked other metrics such as MSE and MAE of the predicted outputs after identifications.

**Fabrication and Alloy compositions**

In the present work, different alloy compositions were obtained by mixing four different base alloys using powder-blown directed energy deposition (DED). The four base alloys used were Nickel-based Inconel 625 (IN625), Cobalt-based Haynes 25 (HA25), Iron-based stainless steel 316L, and a Ni-Co-based MCrAlY thermal spray coat; their compositions are shown in **Table S9**. The alloys were housed in electrostatic powder hoppers (X2W Powder Feeder hoppers, Powder Motion Labs) and powder mass flow was controlled by varying the rotation speed of a metering wheel. Calibration for printing of base alloy fractions was done by measuring the mass flow rate downstream of wye-junction connections at the coaxial annular over 30 s. DED-fabricated superalloy array blocks were stress-relieved under vacuum (900 to 1000 °C for 6-12 h, depending on compositions and build size), cut into plates via wire electrical discharge machining, and ground on both faces using 180 grit SiC paper. Dimples were then drilled in both the additively manufactured plates and a separate hot-rolled wrought IN625 plate (Wr-IN625) and machined to the dimensions shown in **Fig. 1a**.

## Acknowledgements

We acknowledge support from the Defense Advanced Research Projects Agency (DARPA) Multiobjective Engineering and Testing of Alloy Structures (METALS) program project titled "RADICAL: Rapid Array Dimple based Co-design of gradient material and geometry" under cooperative agreement No. HR0011-24-2-0302. This research was supported in part through the computational resources and staff contributions provided for the Quest high performance computing facility at Northwestern University which is jointly supported by the Office of the Provost, the Office for Research, and Northwestern University Information Technology.

## Author contributions

H.D.E. and H.C. designed research; H.C., R.Zhou., R.Zha., Z.C., W.L., J.P.R. conducted DABI experiments, compressive creep experiments, tensile tests, and analyzed data; R.R. fabricated samples; H.D.E., J.C., D.C.D., P.G. proposed research; and H.C. and H.D.E. wrote the paper with the input from all authors.

## Competing interests

The authors declare no competing interests.